%% file: strong_effect_of_cluster_environment_on_protoplanetary_disc_sizes.tex
\documentclass[printer]{aa}

\usepackage{graphicx}
\usepackage{txfonts}
\usepackage{caption}
\usepackage{subcaption}
\usepackage{xcolor}

\newcommand{\noop}[1]{}

\newcommand{\Myr} {\mbox{$~\text{Myr}$}}
\newcommand{\AU} {\mbox{$~\text{AU}$}}
\newcommand{\pc} {\mbox{$~\text{pc}$}}

\newcommand{\pccu} {\mbox{$~\text{pc}^{-3}$}}
\newcommand{\MSun} {\mbox{$~M_{\odot}$}}

\newcommand{\rhoStars} {\mbox{$~\text{stars}~\text{pc}^{-3}$}}

\begin{document} 

   \title{Strong effect of the cluster environment on the size of protoplanetary discs?}

   \author{Kirsten Vincke, Andreas Breslau \and Susanne Pfalzner}

   \institute{Max Planck Institute for Radio Astronomy, Auf dem H\"ugel 69, 53121 Bonn \\
              \email{kvincke@mpifr-bonn.mpg.de}}

   \date{Received ---; accepted ---}


  \abstract
   {Most stars are born in clusters, thus the protoplanetary discs surrounding the newly formed stars might be influenced by this environment. Isolated star-disc encounters have previously been studied, and it
   was shown that very close encounters are necessary to completely destroy discs. However, relatively distant encounters are still able to change the disc size considerably.}
   {We quantify the importance of disc-size reduction that is due to stellar encounters in an entire stellar population.}
   {We modelled young, massive clusters of different densities using the code Nbody6 to determine the statistics of stellar encounter parameters. In a second step, we used these parameters to investigate the
   effect of the environments on the disc size. For this purpose, we performed a numerical experiment with an artificial initial disc size of $10^5\AU$.}
   {We quantify to which degree the disc size is more sensitive to the cluster environment than to the disc mass or frequency. We show that in all investigated clusters a large portion of discs is significantly
   reduced in size. 
   After $5\Myr$, the fraction of discs smaller than 1000 AU in ONC-like clusters with an average number density of \mbox{$\overline{\rho}_{cluster} \sim 60\pccu$}, the fraction of discs smaller than
   $1\,000\AU$  is 65\%, while discs smaller than 100 AU make up 15\%. These fractions increase to 84\% and 39\% for discs in denser clusters like IC 348 \mbox{($\overline{\rho}_{cluster} \sim 500\pccu$)}.
   Even in clusters with a density four times lower than in the ONC \mbox{($\overline{\rho}_{cluster} \sim 15\pccu$),} about $43\%$ of all discs are reduced to sizes below $1\,000\AU$ and roughly $9\%$ to sizes
   below $100\AU$.}
   {For any disc in the ONC that initially was larger than $1\,000\AU$, the probability to be truncated to smaller disc sizes as a result of stellar encounters is quite high. Thus, among other effects, encounters
   are important in shaping discs and potentially forming planetary systems in stellar clusters.}
  

   \authorrunning{Vincke, Breslau \& Pfalzner}
   \maketitle

   \input{introduction}

   \input{method}
   \input{results}
   \input{discussion}
   \input{summary}
   \input{appendix}
   

\bibpunct{(}{)}{;}{a}{}{,} 


\bibliography{strong_effect_of_cluster_environment_on_protoplanetary_disc_sizes}

\Online


\end{document}

%% file: introduction.tex
\section{Introduction}
\label{sec:intro}

Most, if not all stars are initially surrounded by a protoplanetary disc from which a planetary system might eventually form. The observed sizes (= radii) of these discs span from a few $10\AU$ to several
$100\AU$, see \mbox{Table~\ref{tab:observed_disc_sizes}} \citep{Vicente_Alves_2005, Eisner_et_al_2008, McCaughrean_Odell_1996, Andrews_Williams_2007, Andrews_et_al_2009, Andrews_et_al_2010b}. However, the
majority of disc sizes has been found to be in the range of \mbox{$100-200\AU$}. These discs disperse rapidly as a result of internal processes such as viscous torques \citep[e.g.][]{Shu_Adams_Lizano_1987},
turbulent effects \citep{Klahr_Bodenheimer_2003}, and magnetic fields \citep{Balbus_Hawley_2002}. Observations of disc fractions in clusters show that the discs dissipate through internal processes on average at
an age of \mbox{$\leq 2.5\Myr.$} It is currently unclear whether this is valid for young stars in general \citep{Pfalzner_Steinhausen_Menten_2014} and how certain absolute age estimates in clusters are
\citep{Bell_et_al_2013}.

\begin{table}
 \centering
 \caption{Observed disc radii in different cluster environments.}
 \begin{tabular}[b]{lrrr}
  \hline \hline
  Cluster    & Disc radii [AU]              & $N_{\text{sources}}$  & References \\ \hline
  ONC        & $50 - 200\tablefootmark{a}$  & 10                    & (1)        \\
  ONC        & $80 - 217$                   & 39                    & (2)        \\
  ONC        & $27 - 506\tablefootmark{a}$  & 6                     & (3)        \\ 
  Ophiuchus  & $25 - 700$                   & 11\tablefootmark{b}   & (4)        \\
  Ophiuchus  & $14 - 198$                   & 16\tablefootmark{b}   & (5), (6)   \\ 
  Ophiuchus  & $165 - 190$                  & 2                     & (7)        \\ 
  Taurus     & $25 - 600$                   & 12                    & (4)        \\
  Taurus     & $<50 - 100$                  & 4                     & (8)        \\
  \hline
 \end{tabular}
 \tablefoot{Observed clusters (Col.~1) and disc-radius ranges (Col.~2). The number of sources $N_{\text{sources}}$ and the references are given in Cols.~3 and 4. 
 \tablefoottext{a}{Values calculated from given disc diameters.} \tablefoottext{b}{The sources in Cols. (4), (5), and (6) overlap.}}
 \tablebib{(1) \cite{Vicente_Alves_2005}; (2) \cite{Eisner_et_al_2008}; (3) \cite{McCaughrean_Odell_1996}; (4) \cite{Andrews_Williams_2007}; (5) \cite{Andrews_et_al_2009}; (6) \cite{Andrews_et_al_2010b}; 
 (7) \cite{Brinch_Jorgensen_2013}; (8) \cite{Harsono_et_al_2014}.}
 \label{tab:observed_disc_sizes}
\end{table}

There are additional environmental processes in a cluster that, depending on its stellar density, might lead to disc-mass loss, reduction in physical size, or even complete destruction. These processes are
photoevaporation \citep[see e.g.][]{Johnstone_Hollenbach_Bally_1998, Stoerzer_Hollenbach_1999, Scally_Clarke_2002, Matsuyama_Johnstone_Hartmann_2003, Johnstone_et_al_2004, Adams_et_al_2006, 
Alexander_Clarke_Pringle_2006, Ercolano_et_al_2008, Gorti_Hollenbach_2009, Drake_et_al_2009} and stellar fly-bys or gravitational stripping \citep[see e.g.][]{Clarke_Pringle_1993, Scally_Clarke_2002, 
Pfalzner_2004, Adams_et_al_2006, Olczak_Pfalzner_Spurzem_2006, Pfalzner_Olczak_2007, Olczak_Pfalzner_Eckart_2010, Craig_Krumholz_2013, Steinhausen_Pfalzner_2014, Rosotti_et_al_2014}. 

From observations, the sizes of protoplanetary discs are usually determined by fitting the spectral energy distributions (SEDs) to disc models using a truncated power-law \citep{Andrews_Williams_2007}. If
resolved images of the discs can be obtained, the disc size is identified as the radius at which the luminosity drops below a certain value \citep{Vicente_Alves_2005, Odell_1998}. However, it is much more
difficult to observationally constrain the disc size than the disc frequency or disc mass. Therefore, the disc sizes are only determined for a very limited number of stars. As a consequence, it is still a subject
of debate whether the disc size is a function of the mass of the central star \citep[compare e.g.][]{Hillenbrand_et_al_1998, Vicente_Alves_2005, Eisner_et_al_2008}.
The disc frequencies in clusters of similar age \mbox{($\sim 1\Myr$)} reveal that the cluster environment influences the protoplanetary discs. The percentage of stars surrounded by discs decreases with
increasing cluster peak density, dropping from $85\%$ in \mbox{NGC 2024} and $80\%$ in the Orion nebula cluster (ONC) down to $40\%$ in \mbox{NGC 3603} (see also
\mbox{Table~\ref{tab:observed_cluster_properties}}). Because the cluster ages are similar, the higher cluster density leads to a higher degree of disc destruction and thus to a lower disc frequency.

A number of theoretical investigations studying the influence of individual stellar encounters on protoplanetary discs have been performed in the past. For the case of an equal-mass encounter,
\cite{Clarke_Pringle_1993} were the first to study the fate of disc matter after an interaction. They found that a coplanar, prograde encounter at a relative periastron distance of $1.25 r_{\text{peri}}$ material
is removed from the disc down to $0.5 r_{\text{peri}}$. But material is not only removed, it is also re-distributed within the disc, changing the (mass) surface density profile of the disc \citep{Hall_1997}.
As the disc masses of individual stars are observationally better constrained than disc sizes, there has been a considerable amount of theoretical research on the influence of stellar encounters on the disc mass
\citep[e.g.][]{Olczak_Pfalzner_Eckart_2010, Steinhausen_Pfalzner_2014}, but fewer studies of the disc size \citep[e.g.][]{Hall_Clarke_Pringle_1996, Kobayashi_Ida_2001, Breslau_et_al_2014}. Nevertheless, the disc
size is an important parameter because it defines the physical extent of the potentially forming planetary system.

The disc size after an encounter event was investigated for example by \cite{Kobayashi_Ida_2001}. They found that the disc size after an equal-mass encounter is one-third$\mathbf{}$ of the periastron distance. 
\cite{Adams_2010} found a slightly different disc size of b/3 where b is the impact parameter, which is not necessarily equal to the periastron distance. Two other estimates for disc-size upper limits were 
introduced by \cite{de_Juan_Ovelar_et_al_2012}. They derived the disc size assuming that all material outside the Lagrangian point between the two stars is stripped from the  disc. Furthermore, they suggested to
convert the formula for disc-mass loss for parabolic, coplanar encounters reported by \cite{Olczak_Pfalzner_Spurzem_2006} directly into an upper limit for disc truncation, assuming a mass-density distribution
within the disc of \mbox{$\Sigma = r^{-1}$}. 
The few previous studies that considered disc sizes in stellar cluster environments \citep{Adams_et_al_2006, Malmberg_Davies_Heggie_2011, Pfalzner_2013} generalised, for lack of a universal description of disc
sizes, the results obtained for equal-mass encounter by \cite{Kobayashi_Ida_2001} to non-equal mass encounters.

Recently, \cite{Breslau_et_al_2014} showed that a generalisation of this one-third criterion postulated by \cite{Kobayashi_Ida_2001} is not advisable and that the outcome is very sensitive to the mass ratio of
the two stars. They performed simulations of isolated star-disc encounters for a wide range of periastron distances and mass ratios, choosing the highest contrast in surface density to determine the disc sizes
after the encounters. This is similar to the criterion used by observations, and the results by \cite{Breslau_et_al_2014} can easily be applied to cluster simulations. They showed that encounters with large
periastron distances and/or low mass ratios do not remove material from the disc but shift it inwards, reducing the disc size but not the mass. For disc-mass loss closer encounters and/or larger mass-ratios are
needed. This finding demonstrated that the upper disc-size limits of \cite{de_Juan_Ovelar_et_al_2012}, which connect disc-size change directly to disc-mass loss, are thus too high by up to a factor of two.

Recently, \cite{Rosotti_et_al_2014} investigated the effect of encounters on the disc-size distribution in an entire stellar cluster. They performed combined smoothed particle hydrodynamics \mbox{(SPH)/N-body}
simulations of a low-mass cluster \mbox{(100 stars)} allowing hydrodynamical disc spreading. They found that while the masses of the discs are not much affected, the disc sizes are reduced by encounters
within the \mbox{$\sim 0.5\Myr$} of cluster evolution.
 
\begin{table}
 \centering
 \caption{Observed cluster properties.}
 \begin{tabular}[!t]{llll}
  \hline \hline
  Cluster   & $\rho [\pccu]$     & $f_{\text{disc}}$  & References  \\ \hline 
  NGC 2024  & $\approx 10^{3}$   & $85\%$             & (1), (2)    \\
  ONC       & $\approx 10^{4}$   & $80\%$             & (3), (4)    \\
  NGC 3603  & $\approx 10^{5}$   & $40\%$             & (5), (6)    \\ 
  \hline
 \end{tabular}
 \tablefoot{Col.~1 gives the cluster name, $\rho$ is the cluster density (Col.~2), and $f_{\text{disc}}$ the disc fraction (Col.~3). The references for the densities and disc fractions are given in
 Col.~4.}
 \tablebib{(1) \cite{Lada_et_al_1991}; (2) \cite{Haisch_Lada_Lada_2000}; (3) \cite{McCaughrean_Stauffer_1994}; (4) \cite{Lada_et_al_2000}; (5) \cite{Harayama_Eisenhauer_Martins_2008};
 (6) \cite{Stolte_et_al_2004}.}
 \label{tab:observed_cluster_properties}
\end{table}

Here, we wish to analyse numerically how encounters influence the disc size in clusters using a realistic initial mass function (IMF) and the therefore broad spectrum of encounter mass-ratios found in real
clusters. We model clusters of different density to quantify the resulting disc-size dependence. Most importantly, in contrast to previous work, the effects of the encounters are modelled taking into account that
the final disc size is very sensitive to the mass ratio of the two interacting stars. Although external photoevaporation might be as important as stellar encounters when describing disc sizes in clusters, we do
not consider it in this work.

The simulations of the clusters with different densities as well as the disc-size definition are introduced in \mbox{Sect.~\ref{sec:method}}. The choice of initial disc sizes and the classification of encounters
used throughout this work are also discussed there. The most important results are presented in \mbox{Sect.~\ref{sec:results}}, the assumptions applied and their influence on these results are detailed in 
\mbox{Sect.~\ref{sec:discussion_and_conclusion}}. Our findings are briefly summarised in \mbox{Sect.~\ref{sec:summary}}.

%% file: method.tex
\section{Method}
\label{sec:method}

We performed numerical simulations to investigate the influence of star-disc encounters on the disc size within stellar clusters of different densities. This was achieved by performing Nbody6 simulations
\citep{Aarseth_1973, Spurzem_1999, Aarseth_2003} similar to those of \cite{Olczak_Pfalzner_Eckart_2008}, \cite{Olczak_Pfalzner_Eckart_2010}, and \cite{Steinhausen_Pfalzner_2014}. In those simulations, the stellar
encounters in the clusters were tracked and their effect on protoplanetary discs was analysed in the diagnostics using a previously performed parameter study. In contrast to the previous work, we investigate the
disc size rather than the disc mass or frequency in clusters. The disc sizes were obtained using the results from \cite{Breslau_et_al_2014}.

\subsection{Cluster simulations}
\label{sec:cluster_simulations}

The dynamics of the stellar clusters was modelled by Nbody6 simulations taking the cluster density profile of the ONC as a prototype for a massive cluster in the solar neighbourhood. The ONC is the best observed
massive cluster, rendering it possible to compare results of theoretical studies to observational data. It contains about $4\,000$ stars, is \mbox{$\sim 1\Myr$} old, has a half-mass radius of roughly $1\pc$ 
\citep{Hillenbrand_Hartmann_1998}, and extends to \mbox{$\sim 2.5\pc$} \citep{Hillenbrand_Hartmann_1998, Hillenbrand_Carpenter_2000}. \cite{Lada_et_al_2000} found that in the Trapezium, that is, in the inner
$0.3\pc$ of the ONC, about \mbox{$80-85\%$} of the stars are surrounded by a protoplanetary disc, $60\%$ of which having radii smaller than $50\AU$ \citep{Vicente_Alves_2005}.

\cite{Scally_Clarke_McCaughrean_2005} found the density distribution in the ONC to differ from the isothermal approximation \mbox{$\rho_{0} \propto r^{-2}$} in the inner $0.2\pc$ of the cluster.
The surface density there is much flatter with \mbox{$\Sigma \propto r^{-0.5}$}. To obtain this flat profile after $1\Myr$ of simulation time, which is the assumed age of the ONC, \cite{Hillenbrand_1997} and
\cite{Olczak_Pfalzner_Eckart_2010}, for instance, used the three-dimensional initial number density distribution

\begin{equation}
 \rho_{0}(r) = \begin{cases}
  \rho_{0}\left(r/R_{\text{0.2}}\right)^{-2.3}, r \in \left(0, R_{\text{0.2}} \right] & \\
  \rho_{0}\left(r/R_{\text{0.2}}\right)^{-2.0}, r \in \left(R_{\text{0.2}}, R \right] & \\
  0,                                             r \in \left(R, \infty \right], & \\
 \end{cases}
 \label{eq:density_distribution}
\end{equation}

with the radius \mbox{$R_{\text{0.2}}=0.2\pc$}, the cluster radius \mbox{$R=2.5\pc$}, and \mbox{$\rho_{0} = 3.1 \times 10^{3}\rhoStars$}. After $1\Myr$ of cluster evolution this distribution represents today's
number density distribution of the ONC. The stellar positions, masses, and velocities were randomly attributed to the stars. All stars were assumed to be initially single and without any stellar evolution
throughout the simulation. This is standard procedure to reduce the computational cost. Furthermore, the clusters were modelled without gas. The assumptions made and their influence on the outcome is discussed
in \mbox{Sect.~\ref{sec:discussion_and_conclusion}} in detail.

The stellar masses were sampled from the IMF by \cite{Kroupa_2002}

\begin{equation}
 \xi(m) = \begin{cases}
  m^{-1.3}, 0.08 \leq m/\MSun < 0.50 & \\
  m^{-2.3}, 0.50 \leq m/\MSun < 1.00 & \\
  m^{-2.3}, 1.00 \leq m/\MSun < \infty, & \\
  \end{cases}
\end{equation}

with an upper mass limit of $150\MSun$ and randomly attributed to the stars in the cluster. We neglected the effect of initial mass segregation as proposed for the ONC for example by \cite{Bonnell_Davies_1998} and
\cite{Olczak_Spurzem_Henning_2011}. Using an IMF is important, as the effect of unequal-mass encounters onto discs is stronger than for equal-mass encounters
\citep{Olczak_Pfalzner_Spurzem_2006, Moeckel_Bally_2007}.

The clusters were set up to be initially in virial equilibrium: \mbox{$Q = \frac{R_{\text{hm}} \sigma^2}{2GM} = 0.5$}, where $R_{\text{hm}}$ is the half-mass radius, $G$ the gravitational constant,
$M$ the total cluster mass, and $\sigma$ the velocity dispersion. The latter was sampled adopting a Maxwellian distribution.

As we are interested in the effect of clusters of different densities on the disc-size distribution, we set up density-scaled versions of the ONC by keeping the radius fixed at $2.5\pc$ and varying the
number of stars within the cluster. The resulting clusters with $1\,000$, $2\,000$, $8\,000$, $16\,000$, and $32\,000$ stars have average number densities of roughly 1/4, 1/2, 2, 4, and 8 times the density of
the ONC, respectively (see \mbox{Table~\ref{tab:set-up_params}} for the exact number and mass densities). 

For each of these clusters, a whole campaign of similar simulations with different random seeds was performed to minimise the effect of initial stellar properties on the encounter history and thus the
protoplanetary discs. The numbers of simulations $N_{\text{sim}}$ per model are given in \mbox{Table~\ref{tab:set-up_params}}. This approach renders it possible to obtain good statistics of encounters and the
resulting changes in the disc size for environments of different densities.

\begin{table}
 \centering
 \caption{Cluster model set-up.}
 \begin{tabular}[t]{crrrr}
  \hline \hline
  Model  & $N_{\text{stars}}$ & $N_{\text{sim}}$ & $\overline{\rho}_{\text{core}}\tablefootmark{a}$   & $\overline{\rho}_{\text{cluster}}\tablefootmark{b}$  \\ 
         &                    &                  & $[10^3\pccu]$                                      & $[\pccu]$  \\ \hline
  D0     & $1\,000$           & 392              & 1.3                                                & 15.3       \\ 
  D1     & $2\,000$           & 260              & 2.7                                                & 30.6       \\ 
  D2     & $4\,000$           & 264              & 5.3                                                & 61.1       \\ 
  D3     & $8\,000$           & 78               & 10.5                                               & 122.2      \\ 
  D4     & $16\,000$          & 21               & 21.1                                               & 244.5      \\ 
  D5     & $32\,000$          & 14               & 42.1                                               & 489.2      \\
  \hline
 \end{tabular}
 \tablefoot{Model designation (Col.~1), initial number of stars in the cluster $N_{\text{stars}}$ (Col.~2), number of simulations in the campaign $N_{\text{sim}}$ (Col.~3), average number density in the
 cluster core $\overline{\rho}_{\text{core}}$ (Col.~4), and average number density in the whole cluster $\overline{\rho}_{\text{cluster}}$ (Col.~5). \\
 \tablefoottext{a}{The cluster core radius was assumed to be the radius of the Trapezium (\mbox{$R_{\text{core}} = 0.3\pc$}).} \tablefoottext{b}{The cluster radius is $2.5\pc$.}}
 \tablebib{From \cite{Olczak_Pfalzner_Eckart_2010}, \cite{Steinhausen_Pfalzner_2014}.}
 \label{tab:set-up_params}
\end{table}

In each simulation the encounter parameters for star-star interactions were recorded. In a diagnostic step, the disc sizes after each encounter were then calculated with these parameters, as described below in 
\mbox{Sect.~\ref{sec:encounter_simulations}}.

\subsection{Encounter simulations}
\label{sec:encounter_simulations}

\begin{figure*}[t!]
  \centering
  \begin{subfigure}[t]{0.45\textwidth}
    \includegraphics[width=\textwidth]{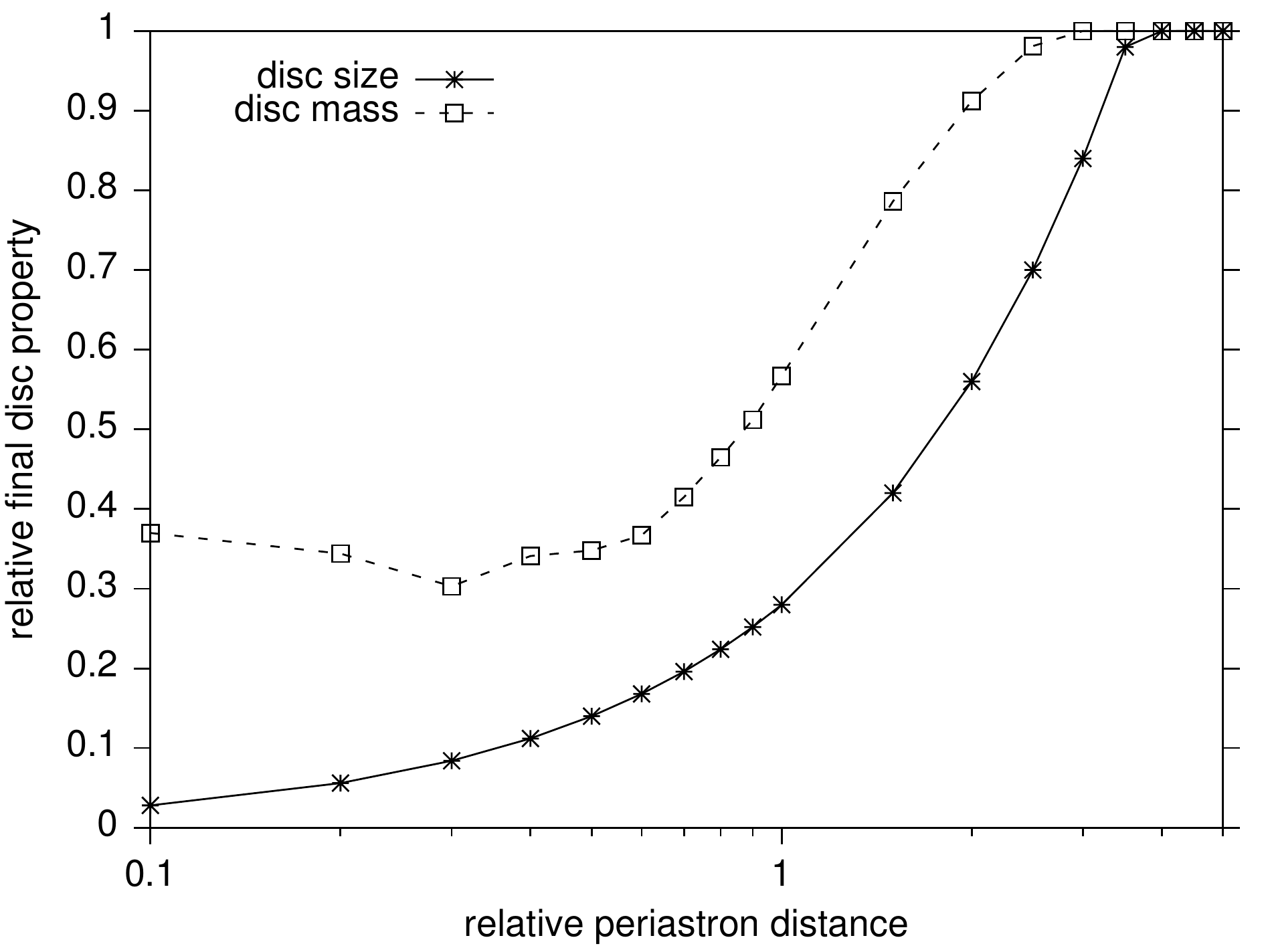}
  \end{subfigure}
  \hfill
  \begin{subfigure}[t]{0.45\textwidth}
    \includegraphics[width=\textwidth]{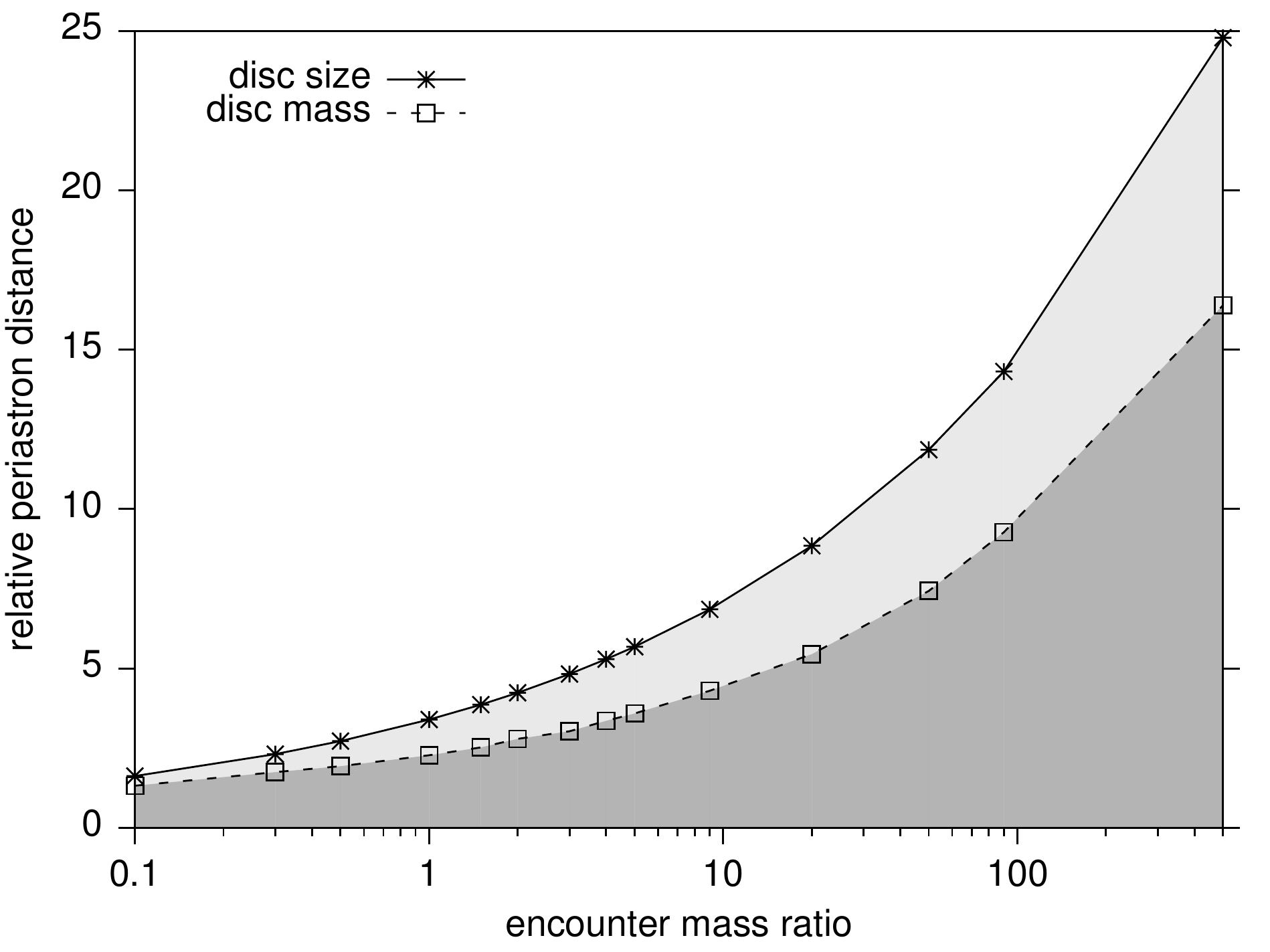}
  \end{subfigure}
  \caption{a) Relative final disc size (asterisks) and disc mass (open squares) after an equal-mass encounter event ($m_{\text{12}}=1$) as a function of the relative periastron distance 
  $r_{\text{p}} = r_{\text{peri}}/r_{\text{previous}}$, with $r_{\text{peri}}$ periastron distance [AU] and $r_{\text{previous}}$ disc size before the encounter [AU].
  b) Parameter pairs (encounter mass ratio $m_{12}$ and relative periastron distance $r_{\text{p}}$) leading to $5\%$ reduction in disc-size (asterisks) and disc mass (open squares). The light and dark grey 
  areas depict the parameter space in which the disc size and mass are reduced.}
  \label{fig:influence}
\end{figure*}

In the diagnostic step the influence of each tracked encounter on the size of the protoplanetary discs was determined. To do
this, we used the results of the parameter study by \cite{Breslau_et_al_2014}.
Since their study was limited to prograde coplanar parabolic (\mbox{eccentricity=1}) encounters, which are the most destructive, the resulting disc sizes should be regarded as lower limits compared to 
hyperbolic and/or inclined encounters. \cite{Breslau_et_al_2014} covered the parameter space for of mass ratios and periastron distances typical for \mbox{star-disc encounters} in cluster environments: 
\mbox{$m_{12} = 0.3-500$} and \mbox{$r_{p} = 0.4-34$}, where \mbox{$m_{12} = m_{2}/m_{1}$} is the ratio between the perturber mass ($m_{2}$) and the mass of the disc-hosting star ($m_{1}$) and
\mbox{$r_{\text{p}} = r_{\text{peri}}/r_{\text{previous}}$} is the periastron distance $r_{\text{peri}}$ normalised to the disc size before the encounter $r_{\text{previous}}$. 

In a diagnostic step, we analysed the cluster simulations described above, using their fit formula 

\begin{equation}
 r_{\text{disc}} = \begin{cases} 0.28 \cdot r_{\text{peri}} \cdot m_{12}^{-0.32}, & \mbox{if } r_{\text{disc}} < r_{\text{previous}} \\ 
                                 r_{\text{previous}},                             & \mbox{if } r_{\text{disc}} \geq r_{\text{previous}},
                   \end{cases}
 \label{eq:breslau_disc_size}
\end{equation}

with $r_{\text{peri}}$ being the periastron distance in AU, to calculate the disc size after each encounter in these clusters.
As \mbox{Eq.~\ref{eq:breslau_disc_size}} shows, the resulting disc size is independent of the initial disc size as long as it is smaller than this initial value. It is basically the strongest encounter that
determines the final disc size.
At the beginning of the simulations, all disc sizes in the clusters were set to a universal value $r_{\text{init}}$ (for details see \mbox{Sect.~\ref{sec:the_initial_disc_size}}). Most
stars in a cluster undergo multiple encounters during the first few Myr. To take those multiple interactions of one and the same star with (different) other cluster members into account, the value of 
$r_{\text{previous}}$ for the star's $i^{th}$ encounter was set to the disc size resulting from the former $(i-1)^{th}$ encounter \mbox{($r_{\text{previous}}^{i} = r_{\text{disc}}^{i-1}$)}. This is an
approximation as the discs might be ``hardened`` by stellar fly-bys or have enough time to spread again between two encounters due to viscosity (see \mbox{Sect.~\ref{sec:discussion_and_conclusion})}. The disc-size
change due to multiple encounters needs to be investigated in
more detail, taking all these effects into account.

Additionally, we assumed that every star in the clusters was initially surrounded by a thin disc of a given size. Although this results in \mbox{disc-disc encounters}, the stellar interactions were treated as
\mbox{star-disc encounters} using \mbox{Eq.~\ref{eq:breslau_disc_size}}. The resulting disc sizes are not expected to differ much from the sizes after disc-disc encounters. For a detailed discussion of the
approximations above, see \mbox{Sect.~\ref{sec:discussion_and_conclusion}} and \cite{Breslau_et_al_2014}.

Before we present the results, we here clarify some definitions.
The code Nbody6 that we used to simulate the clusters was extended to track the stellar encounters and store their properties. For every time step the star with the strongest gravitational influence on a
primary star was identified and its orbit calculated, for additional restrictions and details see \cite{Olczak_et_al_2012}. Most such ''encounters`` are very distant and do not influence the sizes of discs at
all. On the other hand, in extreme cases, stellar encounters cannot only alter the disc size, but destroy them completely. Here, the notion \mbox{``disc-size-reducing encounter``} is used for any encounter that
(a) reduces the disc size by at least $5\%$ relative to its previous disc size and (b) does not destroy the disc \mbox{($r_{\text{disc}} > 10\AU$)}. Consequences of this disc-size reduction limit are discussed in
\mbox{Sect.~\ref{sec:discussion_and_conclusion}}. By contrast, the first encounter that reduces the disc size by at least $5\%$ relative to its previous size and that leads to a final disc size
\mbox{$r_{\text{disc}} < 10\AU$} is denoted as \mbox{''disc-destroying encounter``}. Subsequent, potentially even stronger encounters are not considered in the analysis because the disc is already destroyed. For
these very small discs, effects such as viscous accretion and internal photoevaporation may become important, but they were not considered in this work because of the computational effort. Therefore, all discs
smaller than $10\AU$ are denoted ''destroyed``. The term ''strongest encounter`` in this work describes the encounter with the strongest influence on the disc, which is not necessarily the closest encounter,
because the encounter mass ratio is also factored in.

%% file: results.tex
\section{Results}
\label{sec:results}

Previous studies of protoplanetary discs mostly investigated the effect of star-disc encounters on the disc mass \citep[e.g.][]{Olczak_Pfalzner_Eckart_2010, Steinhausen_Pfalzner_2014}. Before studying this effect
on the \mbox{disc size} in an entire star cluster, we first study the influence of a fly-by on a single star-disc system. Recently, \cite{Rosotti_et_al_2014} noted in their simulations of stellar clusters that
more discs are affected by disc truncation than disc-mass loss\footnote{They used a disc-mass density distribution according to \mbox{$\Sigma \propto r^{2/3}$}}. Here we wish to quantify this higher sensitivity
of the disc size by systematically comparing it to disc-mass loss calculations. We used the results of \cite{Breslau_et_al_2014} and \cite{Steinhausen_Olczak_Pfalzner_2012} to obtain the disc sizes and masses
after stellar encounters. In contrast to the disc size, the disc mass-loss strongly depends on the distribution of material in the disc. Here, we used a disc-mass distribution that follows a power-law:
\mbox{$\Sigma \propto r^{-1}$}.

\subsection{Sensitivity of disc properties}
\label{sec:sensitivity_of_disc_properties}

In \mbox{Fig.~\ref{fig:influence} a)}, the size (asterisks) and mass (open squares) of a disc after an encounter with an equal-mass perturber \mbox{($m_{\text{12}}=1$)} are shown as a function of the relative
periastron distance \mbox{$r_{\text{p}} =r_{\text{peri}}/r_{\text{previous}}$}, where $r_{\text{peri}}$ is the periastron distance and $r_{\text{previous}}$ the disc size before the encounter. The effect of an
encounter on the size is much stronger than on the disc mass. For example, a close grazing encounter, equivalent to \mbox{$r_{\text{p}}=1$}, reduces the disc size to nearly a quarter of its previous size
($28\%$), but it still retains more than half of its mass ($57\%$).

This higher sensitivity of the disc size compared to the disc mass also holds true for encounter partners that differ in mass \mbox{($m_{\text{12}} \neq 1$)}. The shaded areas in
\mbox{Fig.~\ref{fig:influence} b)} depict the parameter space in which the encounters reduce the disc size (asterisks) and disc mass (open squares) by at least $5\%$. At any given mass ratio a reduction of the
disc size by $5\%$ is always achieved already with a much more distant encounter than is necessary for the same reduction in mass. And more importantly, a distant fly-by can affect the disc radius without any
change of the disc mass\footnote{The disc-mass change is also less sensitive than the angular momentum change, see \cite{Olczak_Pfalzner_Spurzem_2006, Pfalzner_Olczak_2007, Steinhausen_Olczak_Pfalzner_2012}.}.
In other words, our data show that, for example, an encounter leading to $5\%$ disc-mass loss can reduce the disc size down to $60\%$ of its previous value. This knowledge can now be used to quantify the
sensitivity of the disc size compared to the disc mass in a whole cluster population.

\subsection{Initial disc size}
\label{sec:the_initial_disc_size}

Most observed discs in clusters have sizes of \mbox{$100-200\AU$}. However, a few protoplanetary discs with radii of a few $100\AU$ were found in the ONC as well. This indicates that a broad spectrum of sizes
exists. It is unclear whether these observed disc sizes are representative for their initial state or whether they changed during the \mbox{$1-2\Myr$} since their formation, nor do we know which role the
environment played in this context. As a consequence, there is no information available about a natal or initial disc-size distribution within the ONC, for instance. 

\begin{figure}[t]
  \centering
  \includegraphics[width=0.45\textwidth]{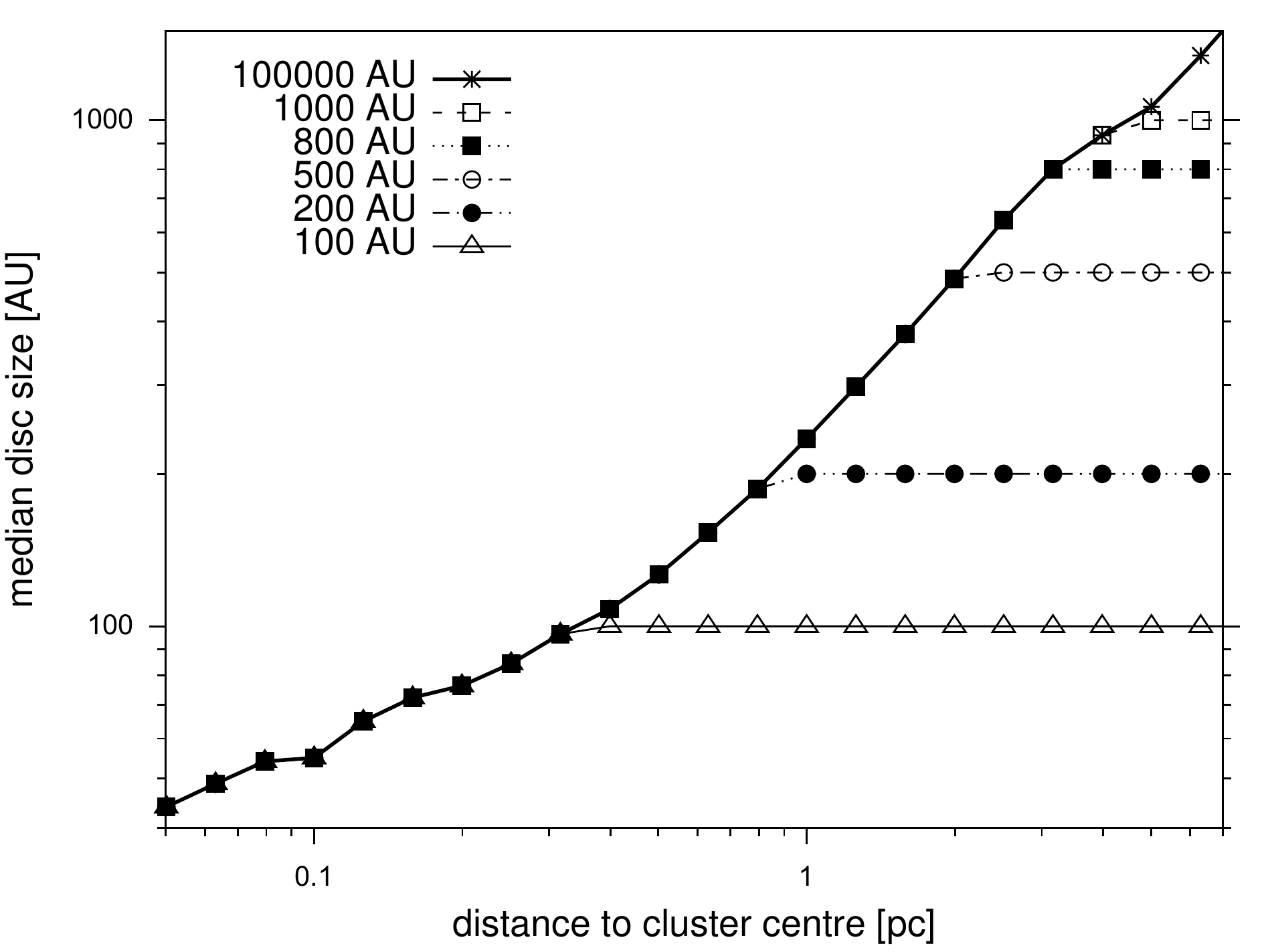}
  \caption{Median disc size as a function of the distance to the cluster centre for model D2 with different initial disc sizes $r_{\text{init}}$: $100\AU$ (open triangles), $200\AU$ (filled circles),
  $500\AU$ (open circle), $800\AU$ (filled squares), $1\,000\AU$ (open squares), and $100\,000\AU$ (asterisks).}
  \label{fig:disc_size_vs_distance_vs_initial_disc_size}
\end{figure}

For this reason, we first investigated how the choice of an initial disc size affects the final disc-size distributions in a cluster. We started our investigation with the \mbox{model cluster D2}, which is
representative for the ONC. All stars in this cluster were initially surrounded by discs of the same size $r_{\text{init}}$. This initial disc size was then varied from $100\AU$ to $100\,000\AU$. Although
most encounters occur in the first $\Myr$ of cluster evolution \citep{Pfalzner_Olczak_Eckart_2006}, the simulation time was chosen to be $5\Myr$ to ensure that all relevant encounters are accounted for.

To obtain an overview of the disc-size distribution in the cluster, the median disc sizes in different radial bins of the cluster were determined. Note that these median disc sizes were determined for each
distance-bin separately, not in a cumulative way. \mbox{Figure~\ref{fig:disc_size_vs_distance_vs_initial_disc_size}} shows these median disc sizes for different initial disc sizes - $100\AU$ (open triangles), 
$200\AU$ (filled circles), $500\AU$ (open circle), $800\AU$ (filled squares), $1\,000\AU$ (open squares), and $100\,000\AU$ (asterisks).

Near the cluster centre the median disc sizes are significantly smaller than in the cluster outskirts. For example, for an initial disc size of $500\AU$, the median disc size at the rim of the cluster core
($<0.3\pc$) is \mbox{$<100\AU,$} but farther out at $2\pc$ it is more than five times as large \mbox{($\sim 500\AU$)}. This trend might be expected because the density is much higher in the inner parts of the
clusters. Thus, more encounters occur there and more discs are reduced in size. Furthermore, the encounters in the inner cluster regions are on average closer, that is, they have lower periastrons 
\citep[see e.g.][]{Scally_Clarke_2001}, which is due to the higher density. Therefore, the discs are smaller than in the less dense cluster outskirts.

Overall, if the discs in the ONC were initially larger than $100\AU$ they are significantly reduced in size by stellar encounters, especially in the cluster core \mbox{($<0.3\pc$)}. Any initial disc size
\mbox{$r_{\text{init}}>100\AU$} leads to the same median disc size distribution in the cluster core. In addition, an initial disc size \mbox{$>1\,000\AU$} was assumed; the final median disc size distribution in
the cluster is the same up to a distance of \mbox{$\sim 4\pc$}, which in the case of the ONC would be the entire cluster \mbox{($R_{\text{cluster}} = 2.5\pc$)}. This finding was then used to set up discs in the
different cluster types to study their influence on the disc-size distribution.

\subsection{Numerical experiment}
\label{sec:the_numerical_experiment}

Building on this finding that the median final disc size in a cluster is largely independent of the initial disc size, we performed a \mbox{numerical experiment} to be able to separate processed from
unprocessed discs. The initial disc sizes in each cluster were set to \mbox{$r_{\text{init}} = 100\,000\AU$} (asterisks in \mbox{Fig.~\ref{fig:disc_size_vs_distance_vs_initial_disc_size})}. Although this is not a
realistic scenario, it allowed us to determine the relevance of encounters in the cluster environments.

This numerical experiment is applied to clusters of different average densities, ranging from \mbox{$15\pccu$ (model D0)} to \mbox{$\sim 500\pccu$ (model D5)}. For the densities of the other models, see 
\mbox{Table~\ref{tab:set-up_params}}. The \mbox{cluster model D2} was denoted \mbox{ONC model} because after 1 Myr of cluster evolution it was representative for today's Orion nebula cluster, see 
\mbox{Sect.~\ref{sec:cluster_simulations}}. Discs that were destroyed \mbox{($r_{\text{disc}} < 10\AU$)} were excluded from the data set and treated separately. If not indicated otherwise, the outcome of the
cluster simulations after $5\Myr$ of evolution is shown.

\subsubsection{Disc-size distributions}
\label{sec:disc_size_distribution}

\begin{figure}[t]
  \centering
  \includegraphics[width=0.45\textwidth]{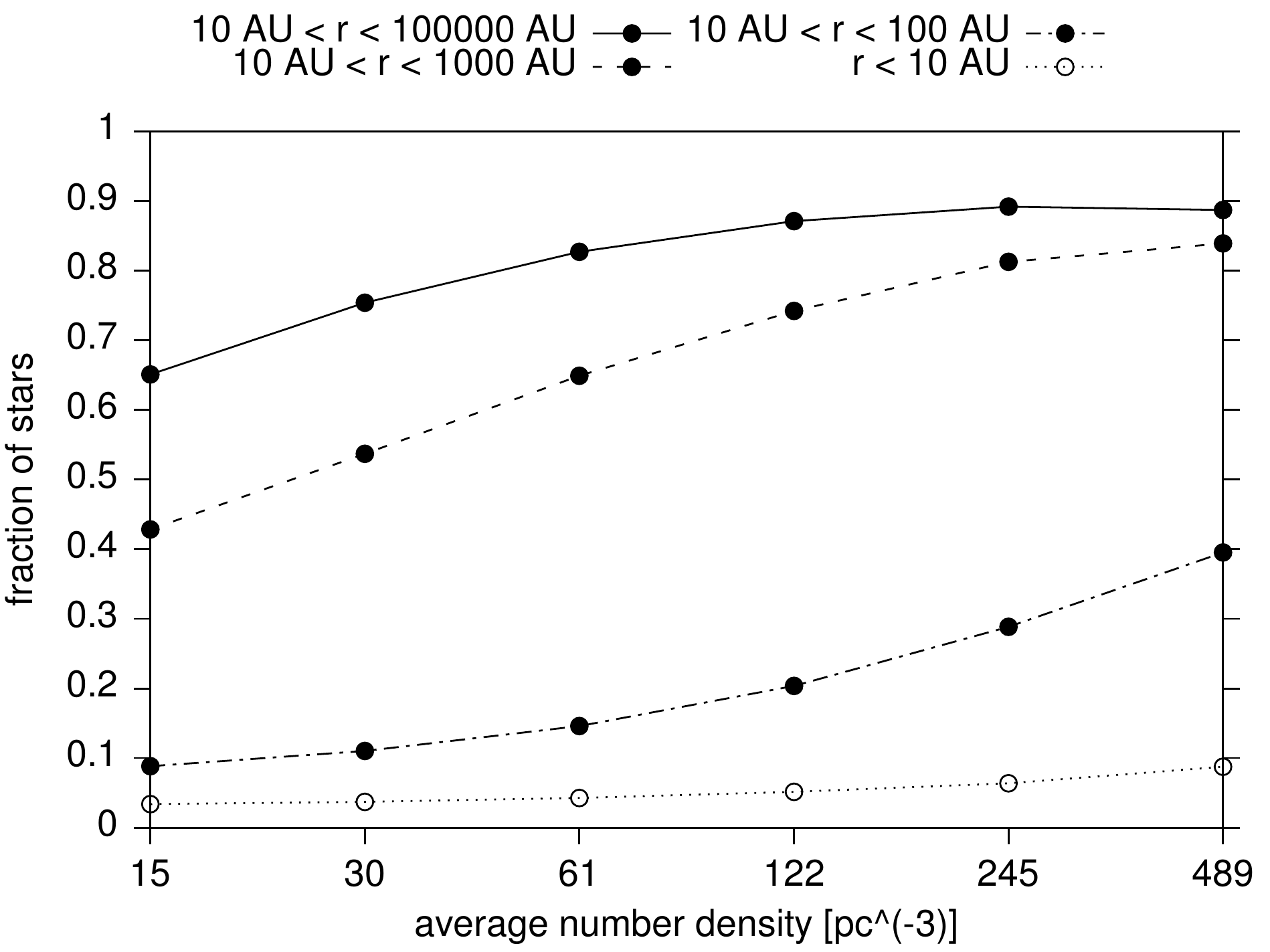}
  \caption{Fraction of stars with destroyed discs (open circles) and whole discs (filled circles) as a function of the average cluster density. Depicted are discs smaller than $100\,000\AU$ (solid line),
  smaller than $1\,000\AU$ (dashed line), and smaller than $100\AU$ (dashed-dotted line) after $5\Myr$.}
  \label{fig:disc_size_smaller_than_XX_cluster_comparison}
\end{figure}

\begin{table}
 \centering
 \caption{Fractions of discs influenced by encounters.}
 \begin{tabular}[t]{crrrrrr}
  \hline \hline
  Model  & $\overline{\rho}_{\text{cluster}}$ & $f_{\text{reduced}}$  & $f_{<1\,000\text{AU}}$  & $f_{<100\text{AU}}$  & $f_{\text{destroyed}}$ \\ 
         & $[\pccu]$                          &                       &                         &                      &                        \\ \hline
  D0     & 15                                 & $65.1\%$              & $42.8\%$                & $8.8\%$              & $3.4\%$                \\ 
  D1     & 30                                 & $75.4\%$              & $53.7\%$                & $11.0\%$             & $3.7\%$                \\ 
  D2     & 61                                 & $82.7\%$              & $64.9\%$                & $14.6\%$             & $4.3\%$                \\ 
  D3     & 122                                & $87.1\%$              & $74.2\%$                & $20.4\%$             & $5.2\%$                \\ 
  D4     & 245                                & $89.2\%$              & $81.3\%$                & $28.8\%$             & $6.4\%$                \\ 
  D5     & 489                                & $88.7\%$              & $83.9\%$                & $39.2\%$             & $8.7\%$                \\
  \hline
 \end{tabular}
 \tablefoot{Fraction of stars with disc-size-reducing encounters \mbox{(Col.~3)}, with discs smaller than $1\,000\AU$ \mbox{(Col.~4)}, with discs smaller than $100\AU$ \mbox{(Col.~5)}, and with
 destroyed discs (Col.~6) in the different cluster models (Col.~1) after $5\Myr$ of cluster evolution. The average number densities of the cluster models are given in \mbox{Col.~2} (rounded values,
 taken from \mbox{Table~\ref{tab:set-up_params}}).}
 \label{tab:fractions_of_dsc_and_dd_enc}
\end{table}

As a first step, we quantified the amount of discs that are influenced by encounters in the different cluster environments. \mbox{Figure~\ref{fig:disc_size_smaller_than_XX_cluster_comparison}} shows the fraction
of stars that undergo at least one disc-size reducing encounter (solid line, filled circles) as a function of the average cluster number density.
The fraction of stars with disc-size-reducing encounters increases, as expected, with cluster density. In the least dense clusters, some stars do not undergo any encounter at all or only very weak ones, that is,
low mass ratios and high periastrons. Here the fraction of stars with disc-size-reducing encounters is \mbox{$\sim 65\%$}. By contrast, in the densest cluster, stellar encounters are very frequent and stronger,
that is, high mass ratios and/or lower periastrons. In this cluster type, nearly all stars \mbox{($\sim 90\%$)} undergo at least one disc-size reducing encounter.

The initial size of the discs was set to $100\,000\AU$, thus, after the first disc-size-reducing encounter it might still be as large as $95\,000\AU$. But most encounters reduce the discs to much lower values and
most discs undergo several disc-size-reducing encounters in the $5\Myr$ of cluster development. Consequently, two questions arise: i) what is the final disc-size distribution? and ii) how many discs are 
destroyed? The dashed and dashed-dotted lines with filled circles in \mbox{Fig.~\ref{fig:disc_size_smaller_than_XX_cluster_comparison}} depict the fractions of discs with a final size smaller than $1\,000\AU$ and
$100\AU$ (see also \mbox{Table~\ref{tab:fractions_of_dsc_and_dd_enc}}).
The fraction of stars with discs smaller than $1\,000\AU$ nearly doubles from low-density clusters ($43\%$) to high-density clusters ($84\%$). Nevertheless, even in environments with densities of $15\pccu$,
stellar encounters significantly reduce the sizes of discs. The slope of the curve flattens towards the high-density cluster end, see \mbox{Fig.~\ref{fig:disc_size_smaller_than_XX_cluster_comparison}}. This
flattening can be explained by studying the fraction of destroyed discs\footnote{A detailed description of the properties of the disc-destroying encounters can be found in the appendix.} (dotted line, open
circles). It amounts to \mbox{$\sim 9\%$} in \mbox{model D5}, which means that in this case, nearly all discs ($97\%$) are either influenced by a disc-size-reducing encounter or are destroyed. This is near
saturation, which is the reason for the flattening of the slope. Furthermore, the fraction of disc-destroying encounters triples from \mbox{$\sim 3\%$} in \mbox{model D0} to \mbox{$\sim 9\%$} in \mbox{model D5}
(see \mbox{Table~\ref{tab:fractions_of_dsc_and_dd_enc}}), whereas the whole discs \mbox{$<1\,000\AU$} only double. This additionally limits the fraction of discs smaller than $1\,000\AU$.

The fraction of stars with final disc sizes \mbox{$<100\AU$} also increases with cluster density (dashed-dotted line). In the high-density cluster model it is four times higher than in the low-density model,
\mbox{$\sim 40\%$} and \mbox{$\sim 10\%$}. In this case, only $19\%$ of all discs in \mbox{model D5} are either reduced to sizes \mbox{$<100\AU$} or destroyed, so there is no flattening of the slope
due to near saturation. 
The fraction of discs \mbox{$<100\AU$} increases much more strongly with cluster density than the fraction of discs \mbox{$<1\,000\AU$}. For example, in the least dense cluster the ratio between discs smaller
than $1\,000\AU$ and discs smaller than $100\AU$ is roughly \mbox{$5:1$}, whereas for the densest cluster it increases to nearly \mbox{$2:1$}. The reason is that the increasing cluster density leads to closer
encounters that in turn reduce many more discs to smaller sizes.
Summarising, the final disc size distribution and the fraction of destroyed discs strongly depends on the average cluster number density.

\subsubsection{Distance to the cluster centre}
\label{sec:the_distance_to_the_cluster_centre}

\begin{figure}[t]
  \centering
  \includegraphics[width=0.45\textwidth]{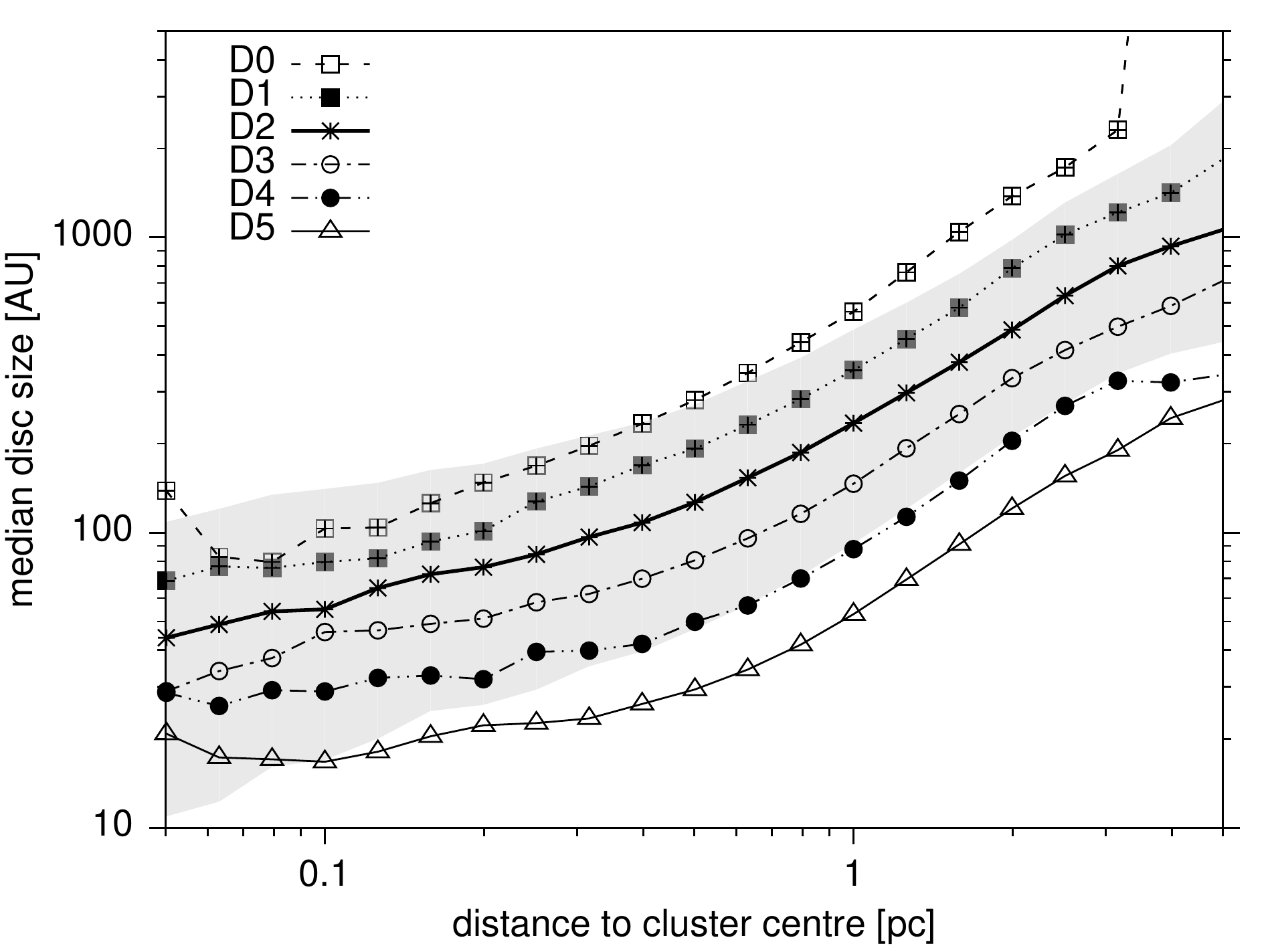}
  \caption{Median disc size in clusters of different densities (rising from D0, open squares, to D5, open triangles) for different distances of the disc-hosting star to the cluster centre (bins). The line with
  asterisks indicates the ONC model, with the grey area containing $50\%$ of all stars with disc sizes around the median disc size. For explanations, see text.}
  \label{fig:disc_size_vs_distance_vs_time}
\end{figure}

We have shown that the final disc sizes strongly depend on the average number density of their cluster environment. Each of these clusters was set up with a density profile according to 
\mbox{Eq.~\ref{eq:density_distribution}}, so they are much denser in the inner regions than in the outskirts. Therefore, the disc sizes should also be a function of the position of the disc-hosting star in the
cluster. In \mbox{Fig.~\ref{fig:disc_size_vs_distance_vs_time}} we compare the median disc sizes as a function of the star's final (after $5\Myr$) distance to the cluster centre in the different clusters. For
comparison, the line with asterisks is the same as in \mbox{Fig.~\ref{fig:disc_size_vs_distance_vs_initial_disc_size}}.

As expected, the median final disc size increases with the distance to the cluster centre for all average cluster densities. Stars that are finally located at a distance of $0.3\pc$ from the cluster centre have
a median disc size of about $200\AU$ in the sparsest cluster \mbox{(D0, open squares)}, whereas in the cluster outskirts it is much larger, for instance \mbox{$\sim 2\,400\AU$} at a distance of $3\pc$. For the
densest clusters \mbox{model D5}, the median disc sizes are much smaller as a result of the elevated rate of close encounters, as described before, but the dependence on the host star position in the cluster is
the same. At $0.3\pc$ the median disc size is about $20\AU,$ whereas at $3\pc$ it is \mbox{$\sim 300\AU$} (open triangles).

To obtain the variation of disc sizes in the clusters, the shaded area in \mbox{Fig.~\ref{fig:disc_size_vs_distance_vs_time}} depicts the disc-size range for $50\%$ of all stars in the ONC model
that have disc sizes of about the median size (asterisks). For example, half of the stars with final distances of $0.3\pc$ to the cluster centre, which represents the border of the Trapezium in the ONC, have
discs with sizes of \mbox{$40-200\AU$}. The median disc size at this distance is $100\AU$. At the cluster radius of the ONC ($2.5\pc$), half of all discs have sizes between \mbox{$300-1\,300\AU$}, the median disc
size is $630\AU$. This means that half of the stars in the ONC model have disc sizes that vary by about a factor of two about the median, but which are still much smaller than the initial disc
size of $100\,000\AU$. The final disc size is a function of the host star position in the cluster, or in other words, not only of the global average cluster density, but also of the local density of its
environment.

\subsubsection{Disc sizes for different stellar types}
\label{sec:disc_sizes_for_different_stellar_types}

\begin{figure}[t]
  \centering
  \includegraphics[width=0.45\textwidth]{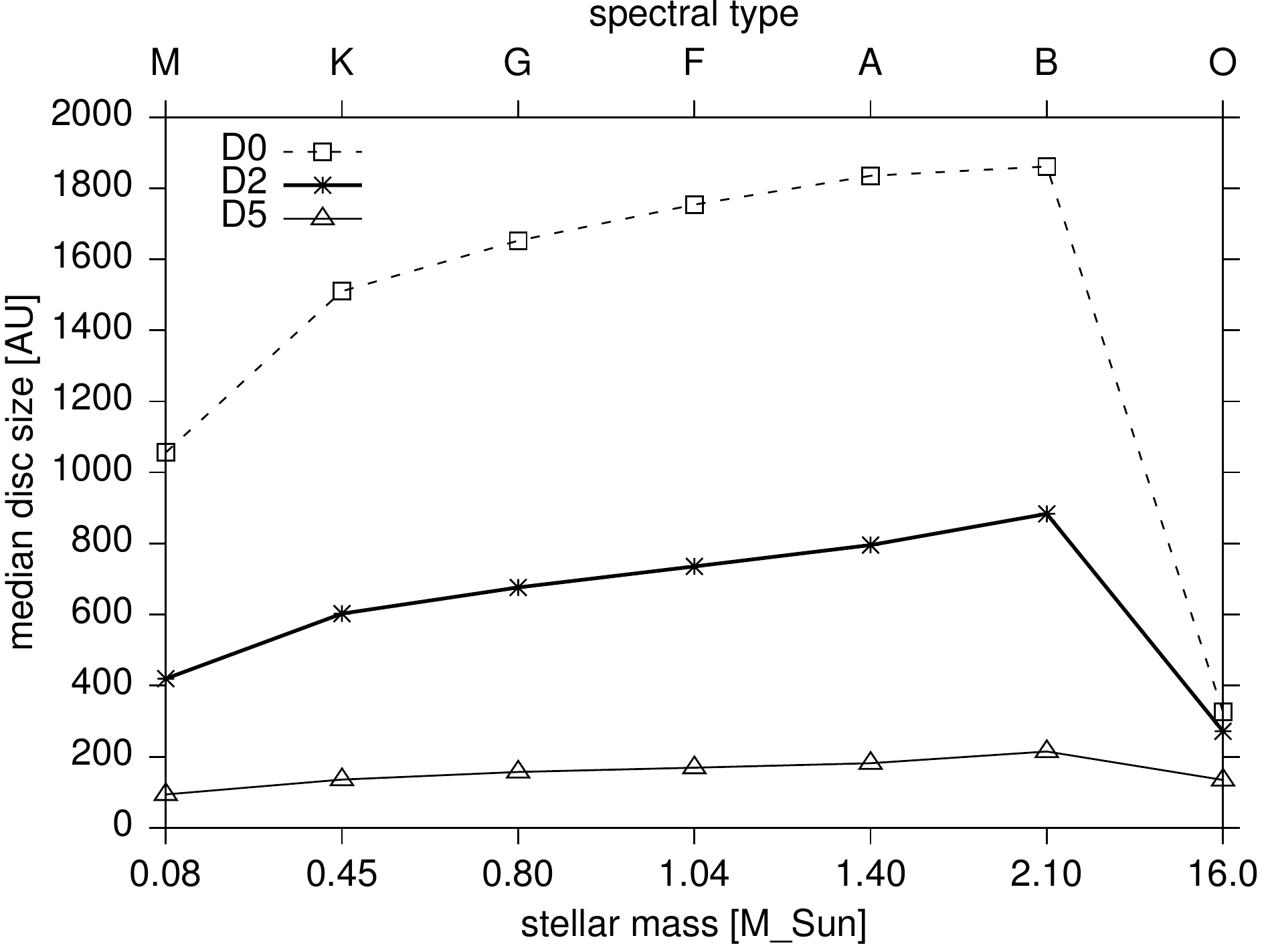}
  \caption{Median disc size in cluster models D0 (open squares), D2 (asterisks), and D5 (open triangles) as a function of the stellar mass [$\MSun$] and spectral type.}
  \label{fig:mean_disc_size_vs_spectral_type}
\end{figure}

As mentioned in \mbox{Sect.~\ref{sec:intro}}, it is still a subject of debate whether the disc size is a function of the stellar mass. We investigated whether encounter-induced disc-size reduction might be
responsible for this potential dependency. \mbox{Figure~\ref{fig:mean_disc_size_vs_spectral_type}} shows the median disc size as a function of stellar mass and spectral type. For clarity, only the results for 
\mbox{models D0 (open squares)}, \mbox{D2 (asterisks)}, and \mbox{D5 (open triangles)} are shown.

The median disc size does indeed depend on the stellar mass. For M- to B-type stars in the ONC model, the median disc size increases by roughly a factor of two, from $420\AU$ to $880\AU$. Around \mbox{O-stars}
the discs are smaller than around all other star types with a median size of $270\AU$. The same trend in median disc sizes can be found for models of lower density (e.g. \mbox{model D0} in
\mbox{Fig.~\ref{fig:mean_disc_size_vs_spectral_type}}). The smaller sizes of discs around \mbox{O-stars} in \mbox{models D0} and D2 can be explained by the encounter frequency, which as a function of the
relative perturber mass has two peaks \citep[see Fig.~4 of][]{Olczak_Pfalzner_Eckart_2010}. One peak lies at low mass-ratios, the other one at high mass-ratios. For \mbox{model D0} the high-mass stars dominate
the encounter history because of gravitational focusing. Therefore, the difference between the median disc sizes of \mbox{M-stars} and \mbox{O-stars} is largest in this model.

In contrast, in \mbox{model D5} the median disc size of \mbox{O-stars} ($130\AU$) is slightly larger than the one of \mbox{M-stars} ($\sim 90\AU$). In this cluster type, the low-mass stars dominate the encounter
history so that \mbox{M-stars} have slightly smaller discs than all other star types - including \mbox{O-stars}. The \mbox{ONC model D2} represents a turning point: the most massive and least massive stars are
equally important as encounter partners. Thus, the absolute difference between the discs around \mbox{M-stars} and \mbox{O-stars} is smaller than in \mbox{model D0}.

All cluster models were set up without primordial mass segregation. Nevertheless, in the $5\Myr$ of cluster evolution, dynamical mass segregation might occur. The final positions of the \mbox{B- and O-stars}
showed no mass segregation in the low-density clusters and only weak mass segregation in the densest models. Therefore, the impact of gravitational focusing on the disc size - especially in the low-density
clusters - is most probably stronger than the effect of dynamical mass segregation.

%% file: discussion.tex
\section{Discussion and conclusion}
\label{sec:discussion_and_conclusion}

Our simulations include several assumptions. Their influence on the results is discussed below.

Most stellar clusters show signs of mass segregation, but whether this is the case for the ONC is still under debate, cf. \cite{Bonnell_Davies_1998, Olczak_Spurzem_Henning_2011, Huff_Stahler_2006}. 
It is currently unclear whether clusters are primordially mass-segregated or not. Here we did not take into account any primordial mass segregation, only dynamical mass segregation during the simulation. This
was done for simplicity because we aimed to study clusters of different densities and their influence on disc sizes. Primordial mass segregation would have increased the initial cluster-core density and altered
the encounter frequency.
Primordial mass segregation affects this type of clusters in two ways: i) the most massive stars undergo many more encounters as a result of their location in the dense cluster centre, and therefore the disc
sizes are smaller around \mbox{O-stars} than in the case without mass-segregation; ii) the stronger gravitational focusing increases the number of encounters. This leads to a somewhat lower disc frequency and
smaller disc sizes, mainly around the \mbox{O-stars}.

All stars in the clusters were assumed to be initially single, neglecting primordial binaries. If these were included, more discs would be truncated to a smaller size or even destroyed by the binary stars,
especially for very close binaries. \cite{Koehler_et_al_2006} found that the binary frequency in the ONC is higher for stars with masses \mbox{$>2\MSun$} than for low-mass stars. Generally, the binary frequency
increases with stellar mass, see \cite{Duchene_Kraus_2013}. In our simulations, primordial binaries would therefore lead to even smaller discs for \mbox{B- and O-stars} than anticipated in 
\mbox{Fig.~\ref{fig:mean_disc_size_vs_spectral_type}}.

The cluster models investigated are representative for young, massive clusters that are still embedded in their natal gas. In the solar neighbourhood the star formation efficiency is relatively low ($30\%$) and 
the gas expulsion drives the expansion of the cluster, while at the same time a large portion of its stars becomes unbound \citep{Pfalzner_Kaczmarek_2013b}. Observations of stellar clusters indicate that after\
\mbox{$1-3\Myr$} the gas is expelled from the cluster and thus the star-formation process ends. We did not consider the effect of gas expulsion. In a follow-up paper, this case of cluster dynamics will be
studied in detail.

We chose an artificially large initial disc of $100\,000\AU$ around every star to determine the relevance of encounters in a cluster environment. Such discs are not physical, but have to be seen as part of the
performed numerical experiment. The resulting median disc sizes can thus not represent any real distribution as found in the ONC, for example, but rather give a first estimate of the influence of encounters on
discs after $5\Myr$ of evolution, especially when considering that other effects are neglected (see below). The surprisingly good match to the observed data shows that the process of disc-size reduction that is
due to stellar encounters is indeed important and should not be neglected. The largest discs in the ONC could have formed even larger and already have been pruned to their current size by dynamical interactions.

All encounters in the clusters were assumed to be prograde, coplanar, and parabolic \mbox{(eccentricity = 1)}. These are the most destructive encounters \citep{Clarke_Pringle_1993, Pfalzner_et_al_2005}.
Especially in very dense clusters, a large portion of encounters can have \mbox{eccentricities > 1}. Such eccentric orbits are less efficient in removing disc mass (\cite{Olczak_Pfalzner_Eckart_2010}). In 
addition, inclined encounters are common in cluster environments. Coplanar encounters lead to smaller disc sizes than their inclined counterparts. Therefore, the cases of parabolic coplanar encounters should be
regarded as lower limits \mbox{(Breslau, in prep.)}.

There are other effects that might influence the disc size. First, we used the formula of \cite{Breslau_et_al_2014} to describe \mbox{star-disc encounters}. It might have been more appropriate to investigate
\mbox{disc-disc encounters}. However, it was shown by \cite{Pfalzner_Umbreit_Henning_2005} that in disc-disc encounters between low-mass discs ($0.01\MSun$) material is exchanged, but it is mainly transferred to
the inner disc regions. Since the disc size is determined using the steepest point in the outer slope of the disc density distribution, this captured disc material is not expected to alter the final size much.

Second, it has been suggested that photoevaporation might be an important mechanism to shrink or completely destroy discs \citep[see e.g.][] {Stoerzer_Hollenbach_1999, Scally_Clarke_2002, Johnstone_et_al_2004,
Adams_et_al_2006, Alexander_Clarke_Pringle_2006, Ercolano_et_al_2008, Gorti_Hollenbach_2009, Drake_et_al_2009}. However, we modelled stellar clusters in the embedded phase, during which stage the presence of the 
natal gas is expected to reduce the effect of photoevaporation. If photoevaporation nevertheless played a role, the presented disc sizes would be even smaller for stars close to the most massive stars, especially
near the cluster core. 

Third, the discs were modelled consisting of mass-less tracer particles and hydrodynamical effects were excluded, for an in-depth discussion see \cite{Breslau_et_al_2014}. \cite{Rosotti_et_al_2014} found that
the hydrodynamical evolution of the discs is important for high viscosity. In this case, the discs spread quite rapidly after an encounter and compensate for the size reduction that is due to encounters. In the
case of low viscosity, it was the closest encounter alone that determined the final disc size. The encounters were particularly important in the dense central cluster areas.
We did not include viscous disc spreading in our simulations. If we had considered this effect, the discs that already underwent their strongest encounter (high mass ratio and/or small periastron
distance) would have been most affected by changes due to disc spreading. If the disc had had time to expand again, the final disc size after a subsequent relatively strong encounter would have been larger than
presented here. To study this effect in more detail, the next natural step would be to include viscous spreading in our type of simulations.

As a result of its sensitivity, the disc size is a suitable tool for quantifying the influence of the stellar environment. One indication that this might be the case comes from the observations by
\cite{Vicente_Alves_2005}. They found that roughly $60\%$ of the Trapezium stars \mbox{($<0.3\pc)$} have discs smaller than $50\AU$. Our simulations indicate that for \mbox{model D2 (ONC)}, the median disc size
of all stars in the cluster core \mbox{($<0.3\pc$)} is roughly $80\AU$. This fits the observed values remarkably well, especially when considering that our simulations only take into account stellar encounters
as a source of disc-size influence. They play an important role in terms of disc-sizes but should be taken together with the other effects discussed above (photoevaporation, viscous disc-spreading, etc.) to
reproduce the actual disc-size distributions as found in stellar clusters today.

The protoplanetary discs may eventually form planetary systems. Their potential physical extent, as shown above, is defined to a high degree by the birth environment. The influence of a stellar cluster on already
formed planetary systems has been studied theoretically in the past \citep[e.g.][]{Craig_Krumholz_2013, Malmberg_Davies_Heggie_2011}. \cite{Craig_Krumholz_2013} found that in dispersing, substructured clusters
planets close to their host stars (few tens of AU distance) are not perturbed by stellar flybys. Compared to the spherical clusters modelled here, the substructures lead to regions of enhanced stellar density and
thus to a higher encounter frequency in certain areas of the cluster. These substructures dissolve in most cases at the latest when the whole cluster disperses. When the substructures are taken as regions of
(transiently) enhanced stellar density, our results support their findings, as only a few per cent of discs are completely destroyed \mbox{($<10\AU$),} while most discs are finally larger than a few tens of AU,
see \mbox{Fig.~\ref{fig:disc_size_smaller_than_XX_cluster_comparison}}. 

%% file: summary.tex
\section{Summary}
\label{sec:summary}

We investigated the influence of the cluster environment on the disc size. Combining Nbody simulations of clusters of different densities with simulations of protoplanetary discs after star-disc encounters,
we found that unlike the disc mass, the disc size can already be changed by relatively distant encounters. We performed a \mbox{numerical experiment}, choosing a very large initial disc size around each
star. The results show that encounters are capable of reducing disc sizes of most stars to values \mbox{$<1\,000\AU$} in all investigated clusters. The resulting sizes agree well with typical disc sizes
found by observations in the ONC today, for example. 
It might be that the cluster environment rather than the initial size determines the final extent of a protoplanetary disc in many cases.\\

%% file: appendix.tex
\begin{appendix}

\section{Properties of disc-destroying encounters}

\begin{figure*}
  \centering
  \begin{subfigure}[t]{0.45\textwidth}
    \includegraphics[width=\textwidth]{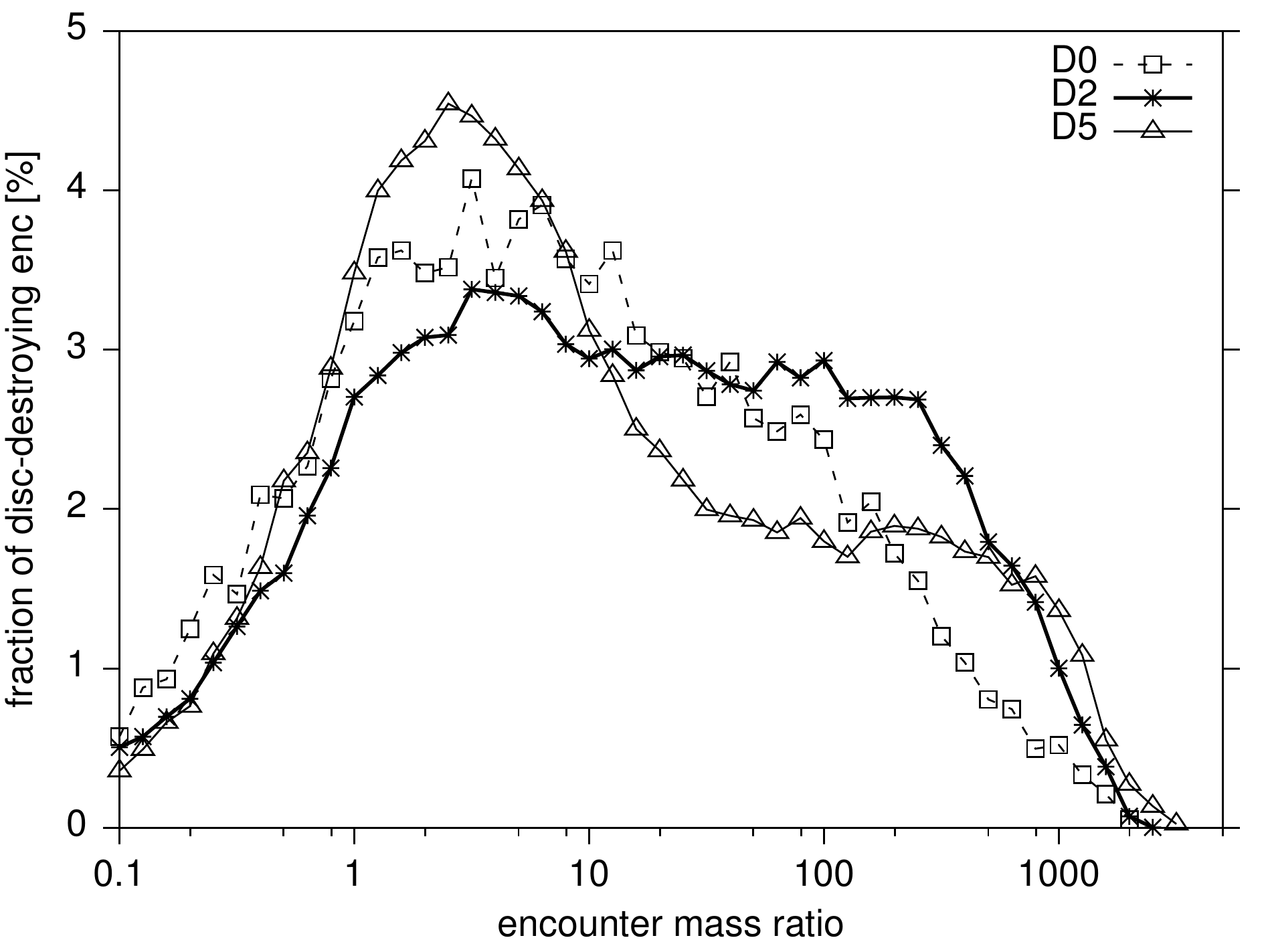}
  \end{subfigure}
  \hfill
  \begin{subfigure}[t]{0.45\textwidth}
    \includegraphics[width=\textwidth]{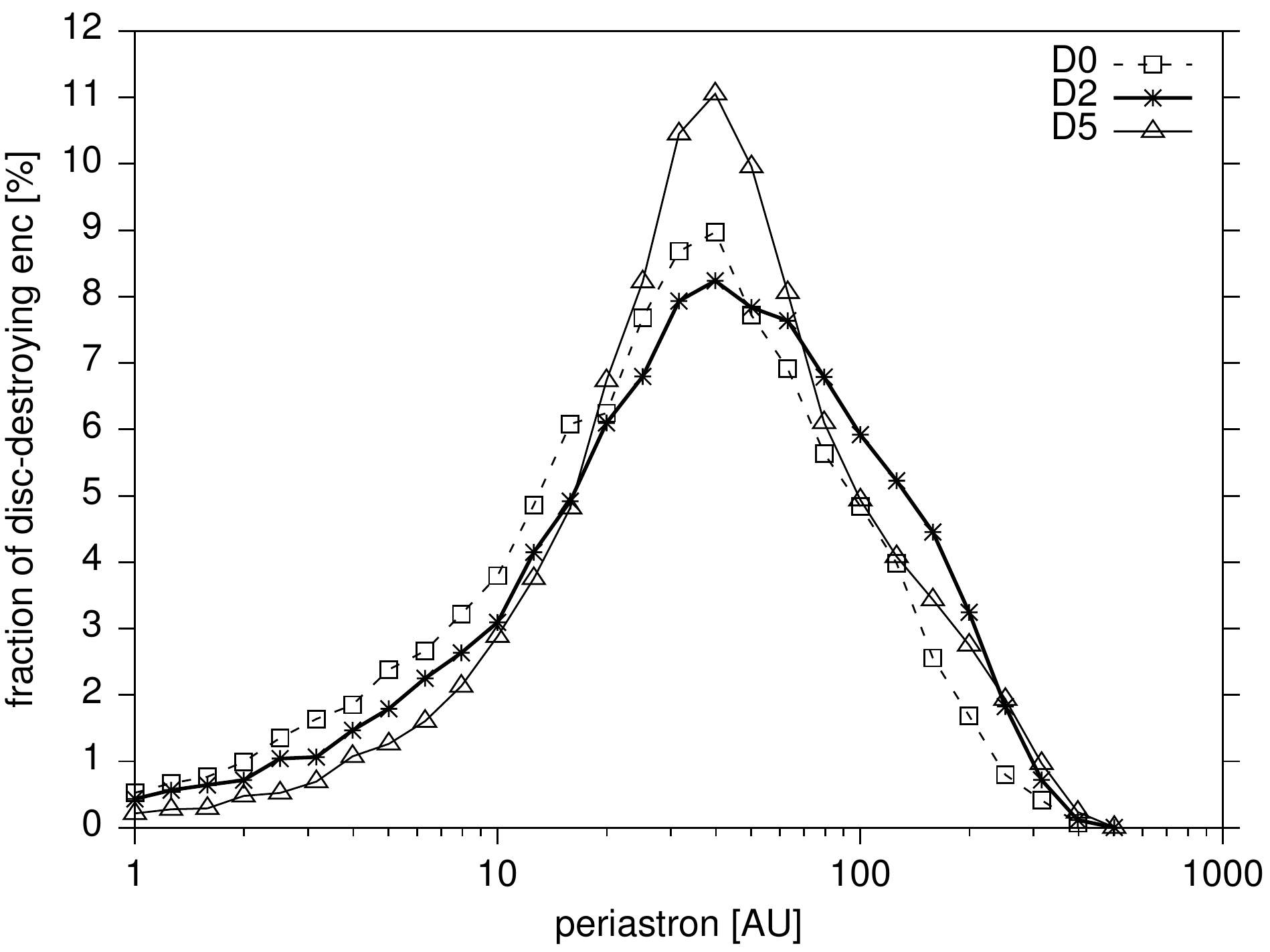}
  \end{subfigure}
  \caption{a) Fraction of disc-destroying encounters as function of the mass ratio and b) fraction of disc-destroying encounters as function of the periastron distance for cluster \mbox{models D0} (open squares),
  D2 (asterisks), and  D5 (open triangles).}
  \label{fig:disc-destroying_enc_properties}
\end{figure*}

\mbox{Figure~\ref{fig:disc-destroying_enc_properties} a)} shows the fraction of disc-destroying encounters as a function of the encounter mass ratio for three cluster \mbox{models (D0, D2, D5)}. 
For the densest cluster model \mbox{(D5, open triangles)}, the encounter mass-ratio distribution has two peaks: one at low mass ratios \mbox{($m_{\text{12}}=3$)} and a second smaller peak at very high mass ratios
\mbox{($m_{\text{12}}=250$)}. So the disc-destroying encounters most probably have quite low mass ratios in this environment. This is in accordance with the results of \cite{Olczak_Pfalzner_Eckart_2010}. They
found that in clusters of higher density than the ONC the low-mass stars dominate the encounter history, leading to lower encounter mass ratios.
They stated that the ONC itself is a turning point, low-mass and high-mass stars similarly contribute to the encounter history. This is reflected in the encounter mass-ratio distribution for the ONC model
(asterisks), which is flat and does not show any sign of a preferred mass ratio for disc-destroying encounters.
For the least dense cluster model \mbox{(D0, open squares)}, one would expect, based on the work of \cite{Olczak_Pfalzner_Eckart_2010}, that high mass-ratio encounters should be largely responsible for disc
destruction, as the most massive stars undergo many encounters due to gravitational focusing. But this is not the case, as \mbox{Fig.~\ref{fig:disc-destroying_enc_properties} a)} shows. The distribution of
disc-destroying encounters has a very broad peak at low mass ratios. 
This might be due to the mass sampling in this work: the masses were randomly sampled from an IMF (see \mbox{Sect.~\ref{sec:method}}) so it is more probable to obtain very massive stars in clusters with higher
membership. In low-mass clusters the most massive star is - on average - less massive than in the high-mass cluster.

\mbox{Figure~\ref{fig:disc-destroying_enc_properties} b)} shows the fraction of disc-destroying encounters as a function of their periastron distance for the same cluster models as above \mbox{(D0, D2, D5)}. For
all cluster types most disc-destroying encounters typically have distances of about \mbox{$30-40\AU$}. The distribution for \mbox{model D2} is slightly broader than those of the other cluster models. The disc
size is determined by both the periastron distance and the mass ratio (see \mbox{Eq.~\ref{eq:breslau_disc_size}}), thus, this is in accordance with the broad spectrum of mass ratios for the disc-destroying
encounters in \mbox{Fig.~\ref{fig:disc-destroying_enc_properties} a)}. If the mass ratio is low, the encounter has to be closer, meaning that the periastron distance needs to be smaller, to destroy a disc than is necessary in a high mass-ratio case.

\section{Location of disc-destroying encounters}
\label{app:location_of_disc-destroying_encounters}

\begin{figure}[t]
  \centering
  \includegraphics[width=0.45\textwidth]{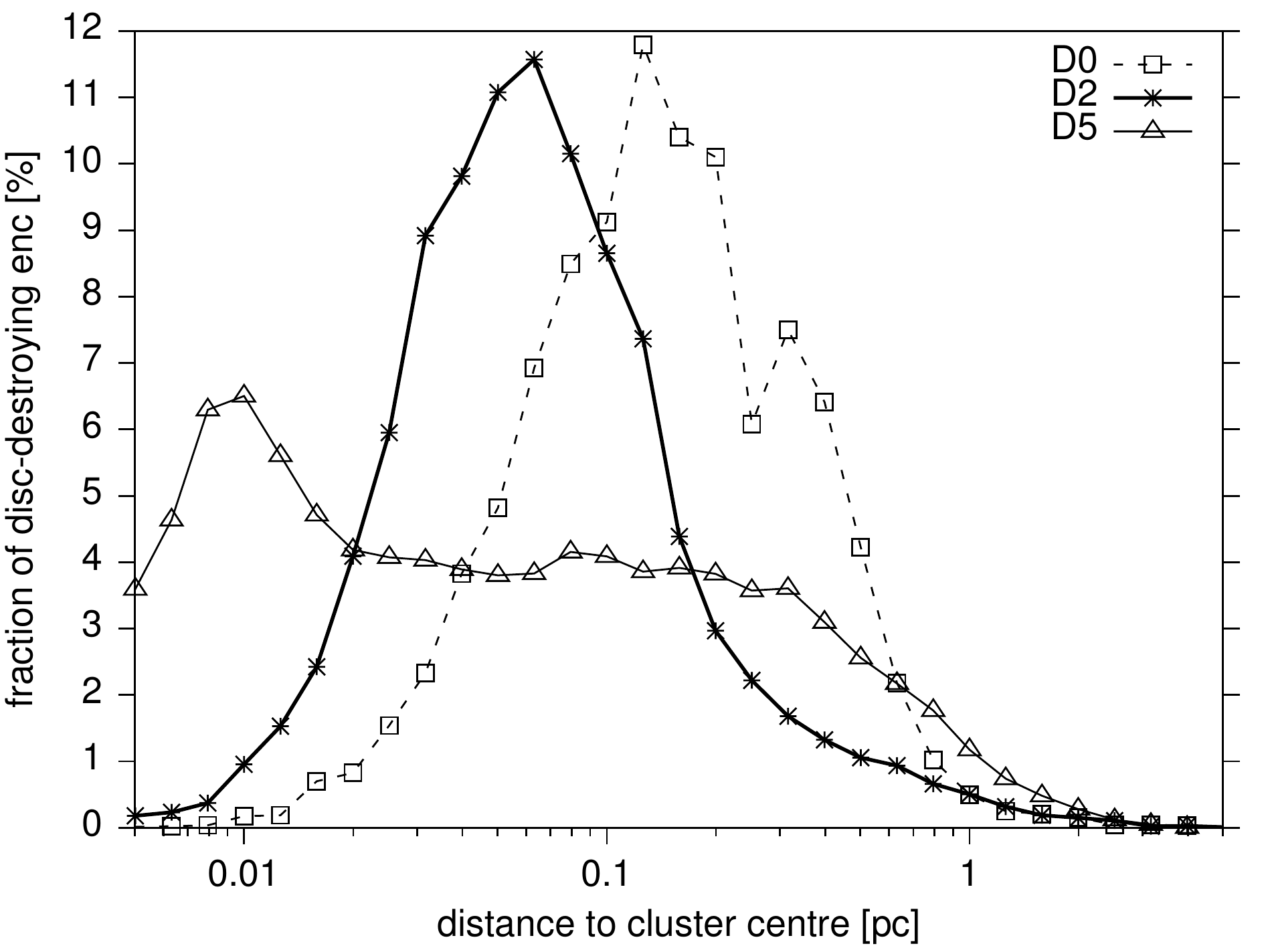}
  \caption{Fraction of disc-destroying encounters as function of distance to the cluster centre where they took place for cluster models D0 (open squares), D2 (asterisks), and
  D5 (open triangles).}
  \label{fig:disc-destroying_enc_vs_dist_to_cluster_center}
\end{figure}

The larger part of discs is influenced by their stellar encounters (\mbox{$65-89\%$}, see \mbox{Table~\ref{tab:fractions_of_dsc_and_dd_enc}}), but is not destroyed. As already mentioned in 
\mbox{Sect.~\ref{sec:disc_size_distribution}}, in the sparsest cluster, about $3\%$ of the discs are destroyed after $5\Myr$; for the densest cluster this fraction increases to \mbox{$\sim 9\%$}. This means that the
fraction of destroyed discs increases with increasing average cluster density. Analogously to the analysis in \mbox{Sect.~\ref{sec:the_distance_to_the_cluster_centre}}, we can now determine whether the
disc-destroying encounters are also a function of the star's position in the cluster, or in other words, a function of the local density.

\mbox{Figure~\ref{fig:disc-destroying_enc_vs_dist_to_cluster_center}} shows the position of the disc-hosting stars at the time of the encounter for three different cluster models (\mbox{D0, open squares,}
\mbox{D2, asterisks}, and \mbox{D5, open triangles}). For the least dense cluster, most disc-destroying encounters take place at a distance of \mbox{$0.1-0.2\pc$} from the cluster centre. In the case of the 
\mbox{ONC model cluster (D2, asterisks)} the whole curve is shifted towards lower distances. As the cluster is much denser than \mbox{model D0}, more destructive encounters occur at such small distances from the
cluster centre. For the densest cluster \mbox{model (D5, open triangles)}, the distribution is entirely different. In this cluster type, the encounters are dominated by low-mass stars, in contrast to the
low-density clusters where the most massive stars dominate the encounter history \citep{Olczak_Pfalzner_Eckart_2010}. As a result, the distribution of locations is much shallower and peaks only in the most inner
part of the cluster \mbox{($\sim 0.01\pc$)}.

\end{appendix}